\documentclass[x11names,dvipsnames,table,xcdraw]{article} 
\usepackage{iclr2024_conference,times}

\usepackage{amsmath,amsfonts,bm}

\def\eqref#1{equation~\ref{#1}}

\def\1{\bm{1}}

\DeclareMathAlphabet{\mathsfit}{\encodingdefault}{\sfdefault}{m}{sl}
\SetMathAlphabet{\mathsfit}{bold}{\encodingdefault}{\sfdefault}{bx}{n}

\usepackage[colorlinks=true,citecolor=brown,urlcolor=gray]{hyperref}
\usepackage{url}
\usepackage{booktabs}
\usepackage{graphicx}
\usepackage{enumitem}
\usepackage{soul}
\usepackage[capitalize]{cleveref}
\usepackage{xcolor}
\usepackage{multirow}
\usepackage{mathtools}
\let\realcite\citep
\renewcommand{\cite}[1]{\ifx.#1.\hl{[?]}\else\realcite{#1}\fi}
\usepackage{longtable}
\usepackage{pifont}
\usepackage{xspace}
\usepackage{wrapfig} 
\usepackage{makecell}

\usepackage{tcolorbox}
\usepackage{enumitem} 

\usepackage{tikz}
\usepackage{mdframed}

\usepackage{pgfplots}
\pgfplotsset{compat=1.18}
\usepackage{pgfplotstable}

\usepackage{subcaption}
\usepackage{caption}

\newcommand\colorLetter[2]{\textcolor{#1}{#2}}
\newcommand\colorLetterScript[2]{\selectfont\fontfamily{pzc}\selectfont \textcolor{#1}{#2}}
\definecolor{RoyalBlue}{rgb}{0.25, 0.41, 0.88} 
\definecolor{CornflowerBlue}{rgb}{0.39, 0.58, 0.93} 
\definecolor{SeaGreen}{rgb}{0.18, 0.55, 0.34} 
\definecolor{LightSeaGreen}{rgb}{0.13, 0.7, 0.67} 
\definecolor{Goldenrod}{rgb}{0.85, 0.65, 0.13} 
\definecolor{Orchid}{rgb}{0.77, 0.33, 0.75} 
\definecolor{MediumOrchid}{rgb}{0.65, 0.2, 0.7} 
\definecolor{Plum}{rgb}{0.8, 0.5, 0.8} 
\definecolor{Thistle}{rgb}{0.75, 0.5, 0.75} 
\definecolor{NotSureBlue}{rgb}{0.396, 0.443, 0.627}

\definecolor{topcolor}{rgb}{1,0.8,0.8}
\definecolor{secondcolor}{rgb}{1,0.87,0.7}
\definecolor{thirdcolor}{rgb}{1,1,0.8}

\newcommand\modeltitle{\textsc{\textbf{\Huge{\colorLetter{RoyalBlue}{T}\colorLetter{CornflowerBlue}{a}\colorLetter{SeaGreen}{n}\colorLetter{LightSeaGreen}{g}\colorLetter{Goldenrod}{o}\colorLetter{Orchid}{F}\colorLetter{MediumOrchid}{l}\colorLetter{Plum}{u}\colorLetter{Thistle}{x}}}}}

\newcommand\model{\textsc{\textbf{{\colorLetter{RoyalBlue}{T}\colorLetter{CornflowerBlue}{a}\colorLetter{SeaGreen}{n}\colorLetter{LightSeaGreen}{g}\colorLetter{Goldenrod}{o}\colorLetter{Orchid}{F}\colorLetter{MediumOrchid}{l}\colorLetter{Plum}{u}\colorLetter{Thistle}{x}}}}}

\newcommand{\stack}[2]{\begin{tabular}{c}#1\\#2\end{tabular}}

\title{\modeltitle{}: Super Fast and Faithful Text to Audio Generation with Flow Matching and Clap-Ranked Preference Optimization}

\author{%
    Chia-Yu Hung$^1$ \quad Navonil Majumder$^1$ \quad Zhifeng Kong$^2$ \quad Ambuj Mehrish$^1$ \\
  \noindent\textbf{Amir Ali Bagherzadeh$^3$ \quad Chuan Li$^3$ \quad Rafael Valle$^2$ \quad Bryan Catanzaro$^2$ \quad Soujanya Poria$^1$} \\[10pt]
    \noindent$^1${Singapore University of Technology and Design (SUTD)} \\
    \noindent$^2${NVIDIA} \\
    \noindent$^3${Lambda Labs} \\[5pt]
    \texttt{\{chiayu\_hung, navonil\_majumder, ambuj\_mehrish, sporia\}@sutd.edu.sg} \\
    \texttt{\{zkong, rafaelvalle, bcatanzaro\}@nvidia.com} \\
    \texttt{\{amirali.zadeh, c\}@lambdal.com} \\
}

\newcommand{\greentick}{\textcolor{teal}{\ding{51}}}
\newcommand{\redcross}{\textcolor{red}{\ding{55}}}

\iclrfinalcopy 
\begin{document}

\maketitle

\begin{figure}[ht]
    \includegraphics[width=\textwidth]{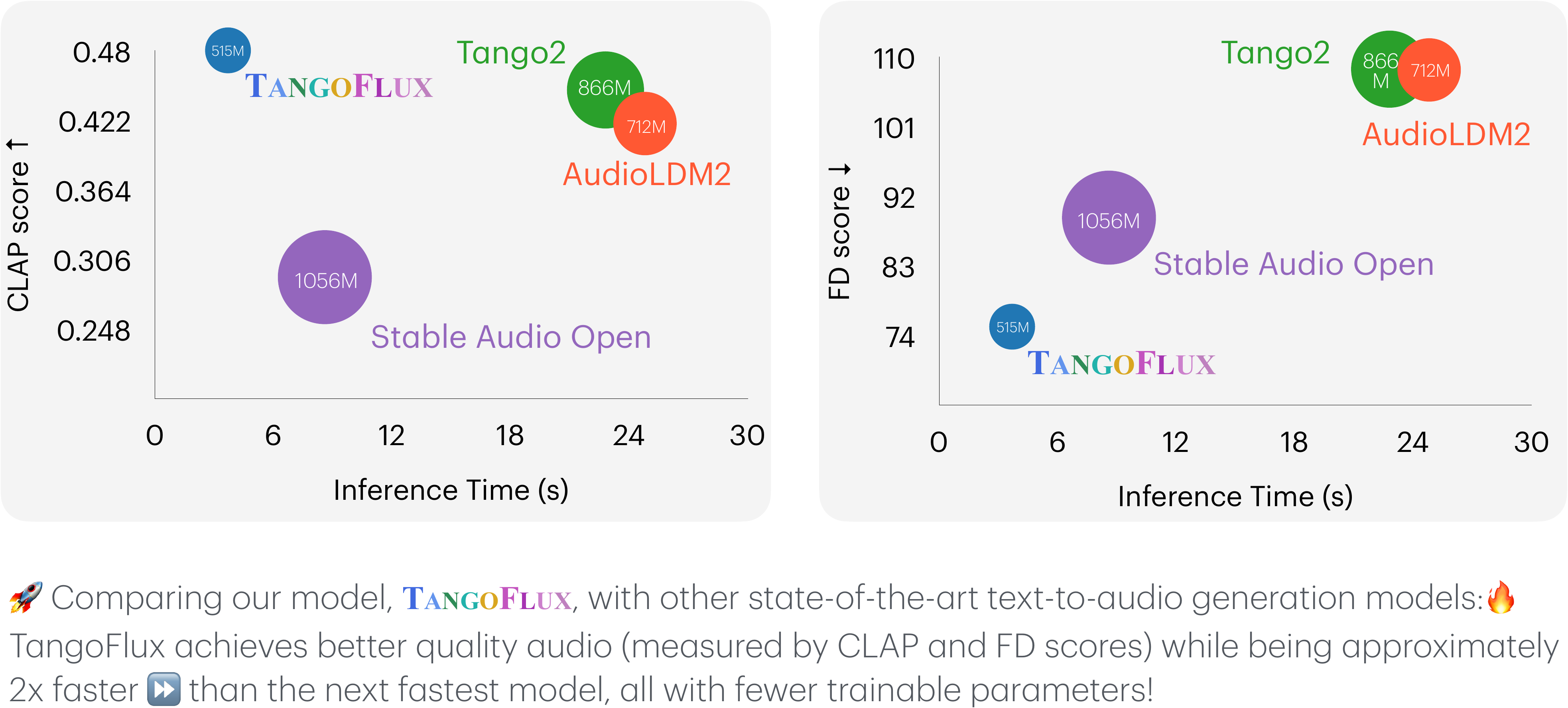}
\end{figure}

\begin{center}

    \textbf{\Large{\model{} \colorLetterScript{SeaGreen}{\textsc{Resources}}}} \\[5pt]
    \begin{tikzpicture}
        \shade[left color=MediumOrchid, right color=Plum]
        (-3,0) .. controls (-2,0.3) and (-1,-0.3) .. (0,0.3)
        .. controls (1,-0.3) and (2,0.3) .. (3,0)
        -- (3,-0.1) -- (-3,-0.1) -- cycle;
    \end{tikzpicture}
    \\[5pt]
    \textbf{\colorLetterScript{RoyalBlue}{Website} → \href{https://tangoflux.github.io}{\colorLetterScript{RoyalBlue}{\texttt{https://tangoflux.github.io}}}} \\[5pt]
    
    \textbf{\colorLetterScript{SeaGreen}{Code Repository} → \href{https://github.com/declare-lab/TangoFlux}{\colorLetterScript{SeaGreen}{\texttt{https://github.com/declare-lab/TangoFlux}}}} \\[5pt]

    \textbf{\colorLetterScript{Goldenrod}{Pretrained Model} → \href{https://huggingface.co/declare-lab/TangoFlux}{\colorLetterScript{Goldenrod}{\texttt{https://huggingface.co/declare-lab/TangoFlux}}}} \\[5pt]
    
    \textbf{\colorLetterScript{Thistle}{Dataset Fork} → \href{https://huggingface.co/datasets/declare-lab/TangoFlux}{\colorLetterScript{Thistle}{\texttt{https://huggingface.co/datasets/declare-lab/CRPO}}}} \\[5pt]

    \textbf{\colorLetterScript{MediumOrchid}{Interactive Demo} → \href{https://huggingface.co/spaces/declare-lab/TangoFlux}{\colorLetterScript{MediumOrchid}{\texttt{https://huggingface.co/spaces/declare-lab/TangoFlux}}}} \\[5pt]
    \begin{tikzpicture}
        \shade[left color=RoyalBlue, right color=Goldenrod]
        (0,0) .. controls (1,0.3) and (2,-0.3) .. (3,0.3)
        .. controls (4,-0.3) and (5,0.3) .. (6,0)
        -- (6,-0.1) -- (0,-0.1) -- cycle;

        \shade[left color=LightSeaGreen, right color=Orchid]
        (6,0) .. controls (7,0.3) and (8,-0.3) .. (9,0.3)
        .. controls (10,-0.3) and (11,0.3) .. (12,0)
        -- (12,-0.1) -- (6,-0.1) -- cycle;
    \end{tikzpicture}
\end{center}

\newcommand{\method}{\textsc{CRPO}}

\definecolor{cycolor}{RGB}{80,24,134}
\newcommand{\CY}[1]{\textcolor{cycolor}{\textbf{CY:} #1}}

\definecolor{nmcolor}{RGB}{194,81,48}
\newcommand{\nm}[1]{\textcolor{nmcolor}{$\ll$\textbf{NM:} #1$\gg$}}

\newcommand{\zk}[1]{\textcolor{blue}{$\ll$\textbf{ZK:} #1$\gg$}}

\begin{abstract}
We introduce \model, an efficient Text-to-Audio (TTA) generative model with 515M parameters, capable of generating up to 30 seconds of 44.1kHz audio in just 3.7 seconds on a A40 GPU. A key challenge in aligning TTA models lies in creating preference pairs, as TTA lacks structured mechanisms like verifiable rewards or gold-standard answers available for Large Language Models (LLMs). To address this, we propose CLAP-Ranked Preference Optimization (\method{}), a novel framework that iteratively generates and optimizes preference data to enhance TTA alignment. We show that the audio preference dataset generated using \method{} outperforms existing alternatives. With this framework, \model{} achieves state-of-the-art performance across both objective and subjective benchmarks.
\end{abstract}

\section{Introduction}

Audio plays a vital role in daily life and creative industries, from enhancing communication and storytelling to enriching experiences in music, sound effects, and podcasts. However, creating high-quality audio, such as foley effects or music compositions, demands significant effort, expertise, and time. Recent advancements in text-to-audio (TTA) generation \cite{majumder2024tango2aligningdiffusionbased,ghosal2023texttoaudiogenerationusinginstructiontuned,liu2023audioldmtexttoaudiogenerationlatent,liu2024audioldm2learningholistic,xue2024auffusionleveragingpowerdiffusion,vyas2023audioboxunifiedaudiogeneration,huang2023makeanaudiotexttoaudiogenerationpromptenhanced,huang2023makeanaudio2temporalenhancedtexttoaudio} offer a transformative approach, enabling the automatic creation of diverse and expressive audio content directly from textual descriptions. This technology holds immense potential to streamline audio production workflows and unlock new possibilities in multimedia content creation. However, many existing models face challenges with controllability, occasionally struggling to fully capture the details in the input prompts, especially when the prompts are complex. This sometimes results in audios that omit certain events or diverges from the user intent. At times, the generated audio may even contain input-adjacent, but unmentioned and unintended, events, that could be characterized as hallucinations.

In contrast, the recent advancements in Large Language Models (LLMs) \cite{ouyang2022traininglanguagemodelsfollow} have been significantly driven by the alignment stage after pre-training and supervised fine-tuning. Alignment often leverages reinforcement learning from human feedback (RLHF) or other reward-based optimization methods to endow the generated outputs with human preferences, ethical considerations, and task-specific requirements \cite{ouyang2022traininglanguagemodelsfollow}. Until recently \cite{majumder2024tango2aligningdiffusionbased}, alignment, that could mitigate the aforementioned issues with audio outputs, has not been a mainstay in TTA model training.

One critical challenge in implementing alignment for TTA lies in the creation of preference pairs. Unlike LLM alignment, where off-the-shelf reward models \cite{lambert2024tulu3pushingfrontiers,lambert2024rewardbenchevaluatingrewardmodels} and human feedback data or verifiable gold answers are available, TTA domain as yet lacks such tooling. 
 For instance, for LLM safety alignment, tools exist for categorizing specific safety risks \cite{inan2023llamaguardllmbasedinputoutput}. 
 
 While audio language models \cite{chu2024qwen2audiotechnicalreport,chu2023qwenaudioadvancinguniversalaudio,tang2024salmonngenerichearingabilities} can take audio inputs and generate textual outputs, they usually produce noisy feedback, unfit for preference pair creation for audio. BATON \cite{liao2024batonaligningtexttoaudiomodel} employs human annotators to assign a binary label 0/1 to each audio sample based on its alignment with a given prompt. However, such labor-intensive manual approach is often impractical at a large scale. 

 To address these issues, we propose CLAP-Ranked Preference Optimization (\method{}), a simple yet effective approach to generate audio preference data and perform preference optimization on rectified flows. As shown in \cref{fig:method}, \method{} consists of iterative cycles of data sampling, generating preference pairs, and performing preference optimization, resembling a self-improvement algorithm. A key aspect of our approach is its ability to evolve by generating its own training dataset, dynamically aligning itself over multiple iterations. We first demonstrate that the CLAP model \cite{laionclap2023} can serve as a proxy reward model for ranking generated audios by alignment with the text description. Using this ranking, we construct an audio preference dataset that post alignment yields superior performance to other static audio preference datasets, such as, BATON and Audio-Alpaca \cite{majumder2024tango2aligningdiffusionbased}.

Many TTA models are trained on proprietary data~\cite{evans2024longformmusicgenerationlatent,evans2024fasttimingconditionedlatentaudio,copet2024simplecontrollablemusicgeneration}, with closed weights and accessible only via private APIs, hindering public use and foundational research. Moreover, the diffusion-based TTA models~\cite{ghosal2023texttoaudiogenerationusinginstructiontuned,majumder2024tango2aligningdiffusionbased,liu2024audioldm2learningholistic} are known to require too many denoising steps for a decent output, consuming much compute and time.

To address this, we introduce \model{}, trained on completely non-proprietary data, achieving \textit{state-of-the-art} performance on benchmarks and out-of-distribution human evaluation, despite its smaller size. \model{} also supports variable-duration audio generation up to 30 seconds with an inference time of ~3.7 seconds on an A40 GPU. This is achieved using a transformer \cite{vaswani2023attentionneed} backbone that undergoes pretraining, fine-tuning, and preference optimization with rectified flow matching training objective---yielding quality audio from much fewer sampling steps.

\textbf{Our contributions}: 
\begin{enumerate}[itemsep=0pt, leftmargin=*, labelwidth=0pt, labelindent=0pt, parsep=0pt, topsep=0pt, label=(\roman*)]
\item We introduce \model{}, a small and fast TTA model based on rectified flow with \textit{state-of-the-art} performance for fully non-proprietary training data. 
\item We propose \method{}, a simple yet effective strategy for dynamically generating audio preference data and aligning rectified flows. By iteratively refining the preference data, \method{} continuously improves itself, outperforming static audio preference datasets. 
\item  We conduct extensive experiments and highlight the importance of each component of \method{} in aligning rectified flows for improving scores on benchmarks.
\item  We plan to release the code and model weights.
\end{enumerate}

\section{Method}

\begin{figure*}
    \centering
    \includegraphics[width=\textwidth]{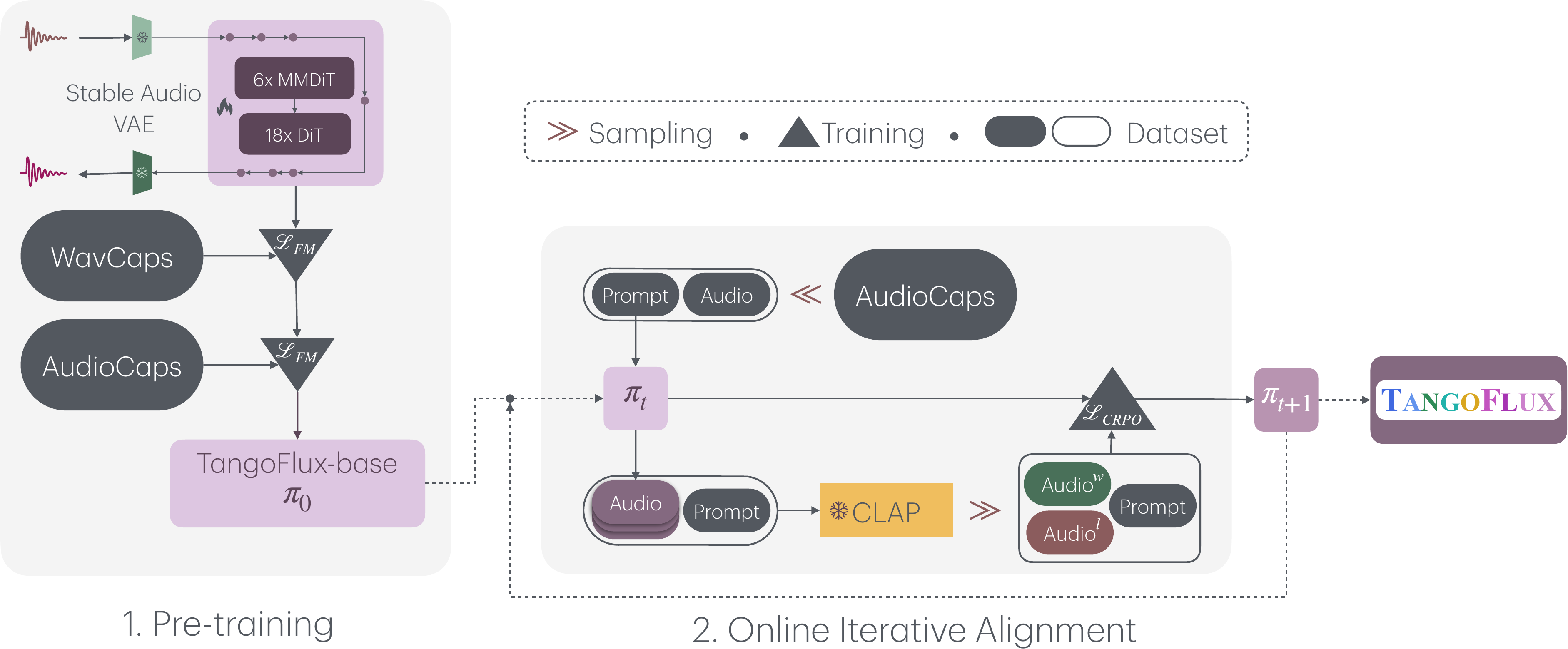}
    \caption{A depiction of the overall training pipeline of \model{}.}
    
    \label{fig:method}

\end{figure*}

\model{} consists of FluxTransformer blocks which are Diffusion Transformer (DiT)~\cite{peebles2023scalablediffusionmodelstransformers} and  Multimodal Diffusion Transformer (MMDiT)~\cite{esser2024scalingrectifiedflowtransformers}, conditioned on textual prompt and duration embedding to generate audio at 44.1kHz up to 30 seconds. 
\model{} learns a rectified flow trajectory to audio latent representation encoded by a variational autoencoder (VAE)~\cite{kingma2022autoencodingvariationalbayes}. As shown in \cref{fig:method}, the training pipeline consists of two stages: pre-training and fine-tuning with alignment. 
\model{} is aligned via \method{} which iteratively generates new synthetic data and constructs preference pairs for preference optimization.

\subsection{Audio Encoding}

We use the VAE from Stable Audio Open~\cite{evans2024stableaudioopen}, which is capable of encoding 44.1kHz stereo audio waveforms into latent representations. Given a stereo audio $X \in \mathbb{R}^{2\times d \times sr}$ with $d$ as the duration and $sr$ as the sampling
rate, the VAE encodes $X$ into a latent representation $Z \in \mathbb{R}^{ L \times C}$, with $L$, $C$  being the latent sequence length and channel size, respectively. The VAE decodes the latent representation $Z$ into the original stereo audio $X$. The entire VAE is kept frozen during \model{} training.

\subsection{Model Conditioning}

To control the generation of audio of varying lengths, we employ (i) text conditioning to control the content of the generated audio and (ii) duration conditioning to dictate the output audio length, up to a maximum of 30 seconds. 

\textbf{Text Conditioning.} We obtain an encoding $c_{text}$ of the given textual description from a pretrained text-encoder. Given the strong performance of FLAN-T5~\cite{chung2022scalinginstructionfinetunedlanguagemodels,raffel2023exploringlimitstransferlearning} as conditioning in text-to-audio generation~\cite{majumder2024tango2aligningdiffusionbased,ghosal2023texttoaudiogenerationusinginstructiontuned}, we select FLAN-T5 as our text encoder.

\textbf{Duration Encoding.} Inspired by the recent works~\cite{evans2024stableaudioopen,evans2024fasttimingconditionedlatentaudio,evans2024longformmusicgenerationlatent}, to generate audios with variable length, we use a small neural network to encode the audio duration into a duration embedding $c_{dur}$ that is concatenated with the text encoding $c_{text}$ and fed into \model{} to control the duration of audio output. 

\subsection{Model Architecture}

Following the recent success of FLUX models in image generation\footnote{\url{https://blackforestlabs.ai/}}, we adopt a hybrid MMDiT and DiT architecture as the backbone for \model{}. While MMDiT blocks demonstrated a strong performance, simplifying some of them into single DiT block improved scalability and parameter efficiency\footnote{\url{https://blog.fal.ai/auraflow/}}. 
These lead us to select a model architecture with $6$ blocks of MMDiT, followed by $18$ blocks of DiT. Each block has $8$ attention heads of $128$ width, totaling a width of $1024$. This setting amounts to 515M parameters.

\subsection{Flow Matching}

Several generative models have been successfully trained under the diffusion framework \cite{ho2020denoisingdiffusionprobabilisticmodels,song2022denoisingdiffusionimplicitmodels,liu2022flowstraightfastlearning}. However, this approach is known to be sensitive to the choice of noise scheduler, which may significantly affect performance. In contrast, the flow matching (FM) framework \cite{lipman2023flowmatchinggenerativemodeling,albergo2023buildingnormalizingflowsstochastic} has been shown to be more robust to the choice of noise scheduler, making it a preferred choice in many applications, including text-to-audio (TTA) and text-to-speech (TTS) tasks \cite{liu2024generativepretrainingspeechflow,le2023voiceboxtextguidedmultilingualuniversal,vyas2023audioboxunifiedaudiogeneration}.

Flow matching builds upon the continuous normalizing flows framework~\cite{onken2021otflowfastaccuratecontinuous}. It generates samples from a target distribution by learning a time-dependent vector field that maps samples from a simple prior distribution (e.g., Gaussian) to a complex target distribution. Prior work in TTA, such as AudioBox \cite{vyas2023audioboxunifiedaudiogeneration} and Voicebox \cite{le2023voiceboxtextguidedmultilingualuniversal}, has predominantly adopted the Optimal Transport conditional path proposed by \cite{lipman2023flowmatchinggenerativemodeling}. However, we utilize rectified flows \cite{liu2022flowstraightfastlearning} instead, which is a straight line path from noise to distribution, corresponding to the shortest path.

\textbf{Rectified Flows.}
Given a latent representation of an audio sample \(x_1\), a noise sample \(x_0 \sim \mathcal{N}(\mathbf{0}, \mathbf{I})\), time-step  $t \in [0,1]$, we can construct a training sample \(x_t\) where the model learns to predict a velocity $v_t = \frac{dx_t}{dt}$ that guides \(x_t\) to \(x_1\). While there exist several methods of constructing transport path \(x_t\) 
, we used rectified flows (RFs) \cite{liu2022flowstraightfastlearning}, in which the forward process are straight paths between target distribution and noise distribution, defined in \cref{eq:flow}. It is empirically shown that rectified flows are more sample efficient and degrade less than other formulations, while consuming fewer sampling steps \cite{esser2024scalingrectifiedflowtransformers}.
The model $u(\mathbf{x}_t,t; \theta)$ directly regresses the ground truth velocity $\mathbf{v}_t$ using the flow matching loss in \cref{eq:flow_loss}.
\begin{flalign}
x_t = (1-t) x_1 + t\tilde{x}_0, v_t = \frac{dx_t}{dt} = \tilde{x}_0 - x_1, \label{eq:flow}\\ 
\mathcal{L}_{\text{FM}} = \mathbb{E}_{x_1,x_0,t}  \left\| u(x_t, t; \theta) - v_t \right\|^2.  \label{eq:flow_loss}
\end{flalign}

\textbf{Inference.} For inference, we sample a noise \(\tilde{x}_0 \sim \mathcal{N}(\mathbf{0}, \mathbf{I})\) and use Euler solver to compute $x_1$, based on the model-predicted velocity $u(\cdot; \theta)$ at each time step $t$.

\subsection{CLAP-Ranked Preference Optimization (CRPO)}

CLAP-Ranked Preference Optimization (CRPO) leverages a text-audio joint-embedding model like CLAP~\cite{laionclap2023} as a proxy reward model to rank the generated audios by similarity with the input description and subsequently construct the preference pairs.

We set $\pi_{0}$ to a pre-trained checkpoint \model{}\texttt{-base} to align. Thereafter, \method{} iteratively aligns checkpoint $\pi_{k} \coloneqq u(\cdot; \theta_k)$ into checkpoint $\pi_{k+1}$, starting from $k = 0$. Each alignment iteration consists of three steps: (i) batched online data generation, (ii) reward estimation and preference dataset creation, and (iii) fine-tuning $\pi_{k}$ into $\pi_{k+1}$ via direct preference optimization. This alignment process allows the model to continuously self-improve by generating and leveraging its own preference data.

This approach of alignment is inspired by a few LLM alignment approaches~\cite{star,kim2024sdpodontusedata,yuan2024selfrewardinglanguagemodels,pang2024iterativereasoningpreferenceoptimization}. However, there are key distinctions to our work: (i) we align rectified flows for audio generation, rather than autoregressive language models; (ii) while LLM alignment benefits from numerous off-the-shelf reward models \cite{lambert2024rewardbenchevaluatingrewardmodels}, which ease the construction of preference datasets based on reward scores, LLM judged outputs, or programmatically verifiable answers, the audio domain lacks such models or method for evaluating audio. We demonstrate that the CLAP model can serve as an effective proxy audio reward model, enabling the creation of preference datasets (see \cref{sec:clap}). Finally, we highlight the necessity of generating online data at every iteration, as iterative optimization on offline data leads to quicker performance saturation and subsequent degradation. 

\subsubsection{CLAP as a Reward Model}
CLAP reward score is calculated as the cosine similarity between textual and audio embeddings encoded by the model. Thus, we assume that CLAP can serve as a reasonable proxy reward model for evaluating audio outputs against the textual description. In \cref{sec:clap}, we demonstrate that using CLAP as a judge to choose the best-of-N inferred policies improves performance in terms of objective metrics.

\subsubsection{Batched Online Data Generation}

To construct a preference dataset at iteration $k$, we first sample a set of prompts $M_k$ from a larger pool $B$. Subsequently, we generate $N$ audios for each prompt $y_i\in M_k$ using $\pi_k$ and use CLAP\footnote{\url{https://huggingface.co/lukewys/laion_clap/blob/main/630k-audioset-best.pt}}~\cite{laionclap2023} to rank those audios by similarity with $y_i$. For each prompt $y_i$, we select the highest-rewarded or -ranking audio $x^w_i$ as the winner and the lowest-rewarded audio $x^l_i$ as the loser, yielding a preference dataset $\mathcal{D}_k = \{(x^w_i,x^l_i,y_i) \mid y_i \in M_k\}$.

\subsubsection{Preference Optimization}

Direct preference optimization (DPO)~\cite{rafailov2024directpreferenceoptimizationlanguage} is shown to be effective at instilling human preferences in LLMs \cite{ouyang2022traininglanguagemodelsfollow}. Consequently, DPO is successfully translated into DPO-Diffusion~\cite{wallace2023diffusionmodelalignmentusing} for alignment of diffusion models.  The DPO-diffusion loss is defined as
\begin{flalign}
   L_{\text{DPO-Diff}} & 
= -\mathbb{E}_{n, \epsilon^w, \epsilon^l} \log \sigma \Big( -\beta \Big[ \nonumber \\
& \| \epsilon^w_n - \epsilon_\theta(x^w_n) \|^2_2  - \| \epsilon^w_n - \epsilon_\text{ref}(x^w_n) \|^2_2 \nonumber \\
& - \big( \| \epsilon^l_n - \epsilon_\theta(x^l_n) \|^2_2 - \| \epsilon^l_n - \epsilon_\text{ref}(x^l_n) \|^2_2 \big) \Big] \Big).
\label{eq:dpo_diffusion_original}
\end{flalign}

 $n\sim U(0,T)$ is a diffusion step among $T$ steps; \( x^l_n \) and \( x^w_n \) represent the losing and winning audios, with \( \epsilon \sim \mathcal{N}(0, \mathbf{I}) \). 
 
Following \citet{esser2024scalingrectifiedflowtransformers}, DPO-Diffusion loss is applicable to rectified flow through the equivalence~\cite{lipman2023flowmatchinggenerativemodeling} between $\epsilon_\theta$ and $u(\cdot; \theta)$,
thereby the noise matching loss terms can be substituted with flow matching terms:
\begin{flalign}
\small
  & L_{\text{DPO-FM}} = -\mathbb{E}_{t\sim \mathcal{U}(0, 1), x^w, x^l} \log \sigma \Big( \nonumber \\
     &-\beta \Big[
        \underbrace{\|u(x^w_t, t; \theta) - v^w_t\|_2^2}_{\text{Winning loss}}
        - \underbrace{\| u(x^l_t, t; \theta) - v^l_t \|_2^2}_{\text{Losing loss}} \nonumber \\
        &- \Big(
            \underbrace{\|u(x^w_t, t; \theta_{\text{ref}}) - v^w_t\|_2^2}_{\text{Winning reference loss}}
            - \underbrace{\|u(x^l_t, t; \theta_{\text{ref}}) - v^l_t \|_2^2}_{\text{Losing reference loss}}
        \Big)
    \Big]
\Big), \label{eq:dpo_rearrange}
 \end{flalign}
where $t$ is the flow matching timestep and \( x^l_t \) and \( x^w_t \) represent losing and winning audio, respectively.

The DPO loss for LLMs models the relative likelihood of the winner and loser responses, allowing minimization of the loss by increasing their margin, even if both log-likelihoods decrease \cite{pal2024smaugfixingfailuremodes}. As DPO optimizes the relative likelihood of the winning responses over the losing ones, not their absolute values, convergence actually requires both likelihoods to decrease despite being counterintuitive \cite{rafailov2024rqlanguagemodel}. The decrease in likelihood does not necessarily decrease performance,
 but required for improvement \cite{rafailov2024scalinglawsrewardmodel}.
 However, in the context of rectified flows, this behavior is less clear due to the challenges in estimating the likelihood of generating samples with classifier-free guidance (CFG). A closer look at $\mathcal{L}_{\text{DPO-FM}}$ (\cref{eq:dpo_rearrange}) reveals that it can similarly be minimized by increasing the margin between the winning and losing losses, even if both losses increase. In \cref{sec:crpo_vsdpo}, we demonstrate that preference optimization of rectified flows via $\mathcal{L}_{\text{DPO-FM}}$ suffer from this phenomenon as well.

 To remedy this, we directly add the winning loss to the optimization objective to prevent \textit{winning loss} from increasing:
\[
\mathcal{L}_{\text{CRPO}} \coloneqq \mathcal{L}_{\text{DPO-FM}}+ \mathcal{L}_{\text{FM}},
\]
where $\mathcal{L}_{\text{FM}}$ is the flow matching loss computed on the winning audio as shown in \cref{eq:flow_loss}. While the DPO loss is effective at improving preference rankings between chosen and rejected audio, relying on it alone can lead to overoptimization. This can distort the semantic and structural fidelity of the winning audio, causing the model’s outputs to drift from the desired distribution. Adding the $\mathcal{L}_{\text{FM}}$ component mitigates this risk by anchoring the model to the high-quality attributes of the chosen data. This regularization stabilizes training and preserves the essential properties of the winning examples, ensuring a balanced and robust optimization process.
Our empirical results demonstrates  $\mathcal{L}_{\text{CRPO}}$ outperform  $\mathcal{L}_{\text{DPO-FM}}$ as shown in \cref{sec:crpo_vsdpo}.

\section{Experiments}
\subsection{Model Training}
We pretrained \model{} on Wavcaps~\cite{mei2023wavcaps} for 80 epochs with the AdamW~\cite{loshchilov2019decoupledweightdecayregularization}, $\beta_1 = 0.9, \beta_2 = 0.95$, a max learning rate of $5\times10^{-4}$. We used a linear learning rate scheduler for $2000$ steps. We used five A40 GPUs with a batch size of $16$ on each device, resulting in an overall batch size of $80$.  After pretraining, \model{} was finetuned on the \textit{AudioCaps} training set for 65 additional epochs. Several works 
find that sampling timesteps $t$ from the middle of its range $[0, 1]$ leads to superior results \cite{hang2024efficientdiffusiontrainingminsnr,kim2024adaptivenonuniformtimestepsampling,karras2022elucidatingdesignspacediffusionbased}, thus, we sampled $t$ from a logit-normal distribution with a mean of $0$ and variance of $1$, following the approach in \cite{esser2024scalingrectifiedflowtransformers}. We name this version as \model{}-\texttt{base}.

During the alignment phase, we used the same optimizer, but an overall batch size of $48$, a maximum learning rate of $10^{-5}$, and a linear warmup of 100 steps. For each iteration of \method{}, we train for $8$ epochs and select the last epoch checkpoint to perform batched online data generation. We performed 5 iterations of \method{} due to the manifestation of performance saturation.

\subsection{Datasets}
\noindent \textbf{Training dataset}. We use complete open source data which consists of approximately 400k  audios from \textit{Wavcaps} \cite{mei2023wavcaps} and 45k audios from the training set of \textit{AudioCaps}. \cite{kim-etal-2019-audiocaps}. Audios shorter than $30$ seconds are padded with silence to 30s. Longer than 30 second audios are center cropped to 30 seconds. Since the audio files are mono, we duplicated the channel to create pseudostereo audios for compatibility with the VAE. 

\noindent \textbf{CRPO dataset.} We initialize the prompt bank as the prompts of \textit{AudioCaps} training set, with a total of ~45k prompts. At the start of each iteration of \method{}, we randomly sample 20k prompts from the prompt bank and generate 5 audios per prompt, and use the CLAP model to construct 20k preference pairs.

\noindent \textbf{Evaluation dataset.} For the main results, we evaluated \model{} on the \textit{AudioCaps} test set, using the same 886-sample split as \cite{majumder2024tango2aligningdiffusionbased}. Objective metrics are reported on this subset. Additionally, we categorized \textit{AudioCaps} prompts using GPT-4 to identify those with multiple distinct events, such as "Birds chirping and thunder strikes," which includes ``sound of birds chirping'' and ``sound of thunder.'' Results on these multi-event captions are reported separately. Subjective evaluation was conducted on an out-of-distribution dataset with 50 challenging prompts.

\subsection{Objective Evaluation}
\textbf{Baselines.} We compare \model{} to three existing strong text-to-audio generation baselines: \texttt{Tango 2}, \texttt{AudioLDM 2}, and \texttt{Stable Audio Open}, including the previous SOTA models. Across the baselines, we use the default recommended classifier free guidance (CFG) scale \cite{ho2022classifierfreediffusionguidance} and the number of steps. For \model{}, we use a CFG scale of 4.5 and 50 steps for inference.  Since \model{} and Stable Audio Open allow variable audio generation length, we set the duration conditioning to 10 seconds and use the first 10 seconds of generated audio to perform the evaluation. We also report the effect of CFG scale in the appendix \ref{app:cfg}.

\textbf{Evaluation metrics.} We evaluate \model{} using both objective and subjective metrics. Following \cite{evans2024fasttimingconditionedlatentaudio}, we report the four objective metrics:
Fréchet Distance (FD$_{\text{openl3}}$) \cite{8682475}, Kullback–Leibler divergence (KL$_{\text{passt}}$) , CLAP$_{\text{score}}$, and Inception Score (IS) \cite{salimans2016improvedtechniquestraininggans}. These metrics allow high-quality audio evaluation up to 48kHz. FD$_{\text{openl3}}$   evaluates the similarity between the statistics of a generated audio set and another reference audio set in the feature space. A low FD$_{\text{openl3}}$ indicates a realistic audio that closely resembles the reference audio.
KL$_{\text{passt}}$ computes the KL divergence over the probabilities of the labels between the generated and the reference audio given the state-of-the-art audio tagger \textbf{PaSST}. A low KL$_{\text{passt}}$ signifies the generated and reference audio share similar semantics tags. 
CLAP$_{\text{score}}$ is a reference-free metric that measures the cosine similarity between the audio and the text prompt. High CLAP$_{\text{score}}$ score denotes the generated audio aligns with the textual prompt. 
IS measures the specificity and coverage of a set of samples. A high IS score represents the diversity in the generated audio.
We use stable-audio-metrics \cite{evans2024fasttimingconditionedlatentaudio} to compute FD$_{\text{openl3}}$, KL$_{\text{passt}}$, CLAP$_{\text{score}}$ and AudioLDM evaluation toolkit \cite{liu2023audioldmtexttoaudiogenerationlatent} to compute \textbf{IS}. Note that we use different CLAP checkpoints to create our preference dataset (\textit{630k-audioset-best}) and to perform evaluation (\textit{630k-audioset-fusion-best})\footnote{\url{https://huggingface.co/lukewys/laion_clap/blob/main/630k-audioset-fusion-best}}. These results are indicated in \Cref{tab:main,tab:multi_event} as CLAP$_{\text{score}}$.

\subsection{Human Evaluation} 

Following prior studies~\cite{ghosal2023texttoaudiogenerationusinginstructiontuned,majumder2024tango2aligningdiffusionbased}, our subjective evaluation covers two key attributes of the generated audio: overall audio quality (OVL) and relevance to the text input (REL). OVL captures the general sound quality, including clarity and naturalness, ignoring the alignment with the input prompt. In contrast, REL quantifies the alignment of the generated audio with the provided text input. At least four annotators rate each audio sample on a scale from 0 (worst) to 100 (best) on both OVL and REL. This evaluation is performed on 50 GPT4o-generated and human-vetted prompts and reported in terms of three metrics: $z$-score, Ranking, and Elo score.
The evaluation instructions, annotators, and metrics are in \cref{app:ood}.

\section{Results}

\subsection{Main Results}

\cref{tab:main} objectively compares \model{} with prior text-to-audio generation models on \textit{AudioCaps}. Performances on the prompts with more than one event, namely \emph{multi-event} prompts, are reported in \cref{tab:multi_event}.
These results suggest that \model{} consistently outperforms the prior works on all objective 
metrics, except \texttt{Tango 2} on KL$_{\text{passt}}$. Interestingly, the margin on CLAP$_{\text{score}}$ between \model{} and baselines is higher for \textit{multi-event} prompts, indicating the superiority of \model{} at grasping complex instructions with multiple events and effectively capturing their nuanced details and relationships in the generated audio.

\begin{table*}[t]
\centering
\resizebox{\textwidth}{!}{
\begin{tabular}{@{}lcccccccc@{}}
\toprule
Model & \#Params.   & Duration & Steps   & FD$_{\text{openl3}}$ $\downarrow$ & KL$_{\text{passt}}$ $\downarrow$ & CLAP$_{\text{score}}$ $\uparrow$ & IS$\uparrow$ &\stack{Inference}{Time (s)}\\ \midrule                        
\texttt{AudioLDM 2}-\texttt{large}   & 712M   & 10 sec & 200   & 108.3    &1.81          &0.419   &7.9 &24.8    \\
\texttt{Stable Audio Open}   & 1056M    & 47 sec & 100   &  89.2   & 2.58  &  0.291   &9.9 & 8.6   \\
\texttt{Tango 2}  & 866M  & 10 sec & 200     & 108.4    & \textbf{1.11}   & 0.447   &9.0 &22.8           \\
\rowcolor{pink!20}
\model{}-\texttt{base} &   \textbf{515M}   & 30 sec & 50 & 80.2& 1.22  & 0.431   &11.7 &\textbf{3.7}  \\
\rowcolor{pink!20}
\model{}    &   \textbf{515M}   & 30 sec & 50 & \textbf{75.1}& 1.15  & \textbf{0.480}   &\textbf{12.2}   & \textbf{3.7}\\
\bottomrule 
\end{tabular}
}
\caption{Comparison of text-to-audio models. Output length represents the duration of the generated audio. Objective metrics include FD$_{\text{openl3}}$ for Fr\'echet Distance, KL$_{\text{passt}}$ for KL divergence, and CLAP$_{\text{score}}$ for alignment. All inferences are performed on the same A40 GPU. We report the trainable parameters in the \#Params column.}
\label{tab:main}

\end{table*}

\begin{table*}[t]
\centering
\resizebox{0.8\textwidth}{!}{
\begin{tabular}{@{}lccccccc@{}}
\toprule
Model                & \#Params.   & Duration   & FD$_{\text{openl3}}$ $\downarrow$ & KL$_{\text{passt}}$ $\downarrow$ & CLAP$_{\text{score}}$ $\uparrow$& IS$\uparrow$ \\ \midrule                        
\texttt{AudioLDM 2}-\texttt{large}   & 712M   & 10 sec   & 107.9    &1.83          &0.415  &7.3     \\
\texttt{Stable Audio Open}   & 1056M    & 47 sec   &  88.5   & 2.67  &  0.286   &9.3 &    \\
\texttt{Tango 2}  & 866M  & 10 sec     & 108.3    & \textbf{1.14}   & 0.452   &8.4 &         \\
\rowcolor{pink!20}
\model{}-\texttt{base} &   \textbf{515M}   & 30 sec & 79.7& 1.23  & 0.438   &10.7 & \\
\rowcolor{pink!20}
\model{}    &   \textbf{515M}   & 30 sec & \textbf{75.2}& 1.20  & \textbf{0.488}   &\textbf{11.1}   & \\  \bottomrule 
\end{tabular}
}
\caption{Comparison of text-to-audio models on multi-event inputs.}
\label{tab:multi_event}
\end{table*}
\subsection{Batched Online Data Generation is Necessary}

\begin{figure}[ht!]
    \centering
\begin{tikzpicture}

    \begin{axis}[
        name=klplot,
        width=7cm,
        height=6cm,
        xlabel={Iteration},
        ylabel={KL Divergence},
        grid=both,
        grid style={dotted, gray!50},
        xtick={1,2,3,4,5},
        xticklabels={1, 2, 3, 4, 5},
        legend style={font=\small, at={(0.5,-0.2)}, anchor=north},
    ]
        \addplot[color=blue, thick, mark=o] coordinates {
            (1,1.18) (2,1.16) (3,1.21) (4,1.16) (5,1.14)
        };
        \addplot[color=red, thick, mark=square*] coordinates {
            (1,1.18) (2,1.19) (3,1.24) (4,1.28) (5,1.35)
        };
    \end{axis}

    \begin{axis}[
        at={(klplot.east)},
        anchor=west,
        xshift=2cm, 
        width=7cm,
        height=6cm,
        xlabel={Iteration},
        ylabel={CLAP$_\text{score}$},
        grid=both,
        grid style={dotted, gray!50},
        xtick={1,2,3,4,5},
        xticklabels={1, 2, 3, 4, 5},
        legend pos=north west,
        legend style={font=\small},
    ]
        \addplot[color=blue, thick, mark=o] coordinates {
            (1,0.453) (2,0.468) (3,0.471) (4,0.48) (5,0.477)
        };
        \addlegendentry{CRPO};
        \addplot[color=red, thick, mark=square*] coordinates {
            (1,0.453) (2,0.467) (3,0.466) (4,0.462) (5,0.448)
        };
        \addlegendentry{CRPO (offline)};
    \end{axis}

\end{tikzpicture}
    \caption{The trajectory of CLAP score and KL divergence across the training iterations. This plot shows the stark difference between online and offline training. Offline training clearly peaks early, by the second iteration, indicated by the peaking CLAP score and increasing KL. In contrast, the CLAP score of online training continues to increase until iteration 4, while the KL divergence has a clear downward trend throughout.
    }
    \label{fig:alignment}
\end{figure}

In \cref{fig:alignment}, we present the results of five training iterations of \method{}, both with and without generating new data at each iteration. Our findings suggest that training on the same dataset over multiple iterations leads to quick performance saturation and eventual degradation. Specifically, for offline \method{}, the CLAP score decreases after the second iteration, while the KL increases significantly. By the final iteration, the performance degradation is evident from CLAP score and KL being worse than first iteration, emphasizing the limitations of offline data. In contrast, the online \method{} with data generation before each iteration outperforms the offline \method{} in terms of CLAP score and KL.

This performance degradation could be ascribed to reward over-optimization~\cite{rafailov2024scalinglawsrewardmodel,gao2022scalinglawsrewardmodel}. \citet{kim2024sdpodontusedata} showed that the reference model serves as a regularizer in DPO training for language models. Several iterations of updating the reference model with the same dataset thus may hamper the due regularization of the loss. In \cref{sec:crpo_vsdpo}, we show the paradoxical performance degradation with loss minimization, indicating over-optimization.

\begin{table}[t]
\centering
\small
\setlength{\tabcolsep}{3pt}
\begin{tabular}{@{}lccc@{}}
\toprule
Dataset        & FD$_{\text{openl3}}$ $\downarrow$ & KL$_{\text{passt}}$ $\downarrow$ & CLAP$_{\text{score}}$ $\uparrow$ \\ \midrule
BATON          & 80.5                              & 1.20                              & 0.437                             \\
Audio Alpaca   & 80.0                              & 1.20                              & 0.448                             \\
\rowcolor{pink!20}
\method{}      & \textbf{79.1}                     & \textbf{1.18}                     & \textbf{0.453}                    \\
\bottomrule
\end{tabular}
\caption{Comparison of \model{} checkpoints aligned with three preference datasets. FD$_{\text{openl3}}$ $\coloneqq$ Fr\'echet Distance and KL$_{\text{passt}}$ $\coloneqq$ KL divergence.}
\label{tab:baton}
\end{table}

\subsection{CLAP as Reward Model}\label{sec:clap}

To validate CLAP as a proxy reward model for evaluating audio output, we further evaluate \model{} under a CLAP-driven Best-of-$N$ policy, where $N \in \{1,5,10,15\}$. We use CLAP \textit{630k-audioset-best.pt} checkpoint to rank the generated audios. The results in \cref{tab:bon} suggest that increasing $N$ yield better CLAP$_{\text{score}}$ and KL$_{\text{passt}}$ while FD$_{\text{openl3}}$ remains about the same. This indicates that the CLAP can identify well-aligned audio outputs that better represent the textual descriptions, without compromising diversity or quality, as implied by the lower KL$_{\text{passt}}$ and similar FD$_{\text{openl3}}$.

\begin{figure}[htbp]
    \centering
\begin{tikzpicture}
    \begin{axis}[
        width=10cm,
        height=7cm,
        xlabel={Iteration},
        ylabel={Loss},
        grid=both,
        grid style={dotted, gray!50},
        xtick={1,2,3,4,5},
        xticklabels={1, 2, 3, 4, 5},
        legend style={font=\small, at={(0.5, -0.2)}, anchor=north, legend columns=2},
        cycle list name=color list
    ]

        \addplot[color=blue, thick, mark=o] coordinates {
            (1,1.5207) (2,1.5229) (3,1.5159) (4,1.52476) (5,1.5389)
        };
        \addlegendentry{Winning Loss (CRPO)};

        \addplot[color=red, thick, mark=square*] coordinates {
            (1,1.5209) (2,1.5242) (3,1.5205) (4,1.53379) (5,1.5551)
        };
        \addlegendentry{Losing Loss (CRPO)};

        \addplot[color=green!60!black, thick, mark=triangle*] coordinates {
            (1,1.5204) (2,1.5255) (3,1.53821) (4,1.569) (5,1.6273)
        };
        \addlegendentry{Winning Loss (DPO-Diff)};

        \addplot[color=purple, thick, mark=star] coordinates {
            (1,1.5211) (2,1.5285) (3,1.54542) (4,1.5837) (5,1.6572)
        };
        \addlegendentry{Losing Loss (DPO-Diff)};

    \end{axis}
\end{tikzpicture}
    \caption{Winning and Losing losses of $\mathcal{L}_{\text{DPO-FM}}$ and $\mathcal{L}_{\text{CRPO}}$ at each iteration. Winning and Losing losses increase each iteration, as well as their margin.}
    \label{fig:crpo_dpo}
\end{figure}

\subsection{\method{} surpasses Static Audio Preference Datasets}
To show the superiority of \method{}, we compare its performance with two other static audio preference datasets: Audio-Alpaca \cite{majumder2024tango2aligningdiffusionbased} and BATON \cite{liao2024batonaligningtexttoaudiomodel} (see \cref{app:baton} for details).



We apply preference optimization to \model{}-\texttt{base}, lasting only one iteration since Audio-Alpaca and BATON are fixed datasets. \cref{tab:baton} reports objective metrics FD${\text{openl3}}$, KL${\text{passt}}$, and CLAP${\text{score}}$, demonstrating that preference optimization with the CRPO dataset outperforms the other two audio preference datasets across all metrics. Despite its simplicity, \method{} proves highly effective for constructing audio preference datasets for optimization. 

\begin{table}[h]
\small
\centering
\resizebox{0.65\linewidth}{!}{
\begin{tabular}{@{}lcccccc@{}}
\toprule
Model & N & FD$_{\text{openl3}}$ $\downarrow$ & KL$_{\text{passt}}$ $\downarrow$ & CLAP$_{\text{score}}$ $\uparrow$ \\ \midrule                        
\multirow{4}{*}{\textbf{\model{}}} & 1 & 75.0 & 1.15 & 0.480 \\
                                   & 5 & 74.3 & 1.14 & 0.494 \\
                                   & 10 & 75.8 & 1.08 & 0.499 \\
                                   & 15 & 75.1 & 1.11 & 0.502 \\ \midrule
\multirow{4}{*}{\texttt{Tango 2}}  & 1 & 108.4 & 1.11 & 0.447 \\
                                   & 5 & 108.8 & 1.05 & 0.467 \\
                                   & 10 & 108.4 & 1.08 & 0.474 \\
                                   & 15 & 108.7 & 1.06 & 0.473 \\
\bottomrule 
\end{tabular}
}
\caption{Best-of-$N$ FD, KL, and CLAP scores.}
\label{tab:bon}
\end{table}


\subsection{$\mathcal{L}_{\text{CRPO}}$ vs $\mathcal{L}_{\text{DPO-FM}}$}\label{sec:crpo_vsdpo}

 To study the relationship between the winning and losing losses of $\mathcal{L}_{\text{CRPO}}$ and $\mathcal{L}_{\text{DPO-FM}}$ (see \cref{eq:dpo_rearrange}), we calculate the average winning and losing losses of the final checkpoint (epoch 8) of each iteration on the training set. The losses are plotted in \cref{fig:crpo_dpo}. Simultaneously in \cref{fig:crpovsdpo}, we present the benchmark performances of the checkpoints by $\mathcal{L}_{\text{CRPO}}$ and $\mathcal{L}_{\text{DPO-FM}}$ on \textit{AudioCaps} training set. Here, we only use fixed preference data by \model{}-\texttt{base}.

 
\begin{figure}[htbp]
    \centering
    \begin{subfigure}[t]{0.32\textwidth}
        \centering
        \begin{tikzpicture}
        \begin{axis}[
            width=1.2\textwidth,
            height=1.0\textwidth,
            xlabel={Iteration},
            xtick={1,2,3,4,5},
            ylabel={},
            grid=both,
            grid style={dotted, gray!50},
            mark options={solid},
            ticklabel style={font=\small}
        ]
        \addplot[color=blue, thick, mark=o] coordinates {
            (1, 0.450) (2, 0.460) (3, 0.460) (4, 0.447) (5, 0.418)
        };
        \addplot[color=red, thick, mark=o] coordinates {
            (1, 0.453) (2, 0.467) (3, 0.466) (4, 0.462) (5, 0.448)
        };
        \end{axis}
        \end{tikzpicture}
        \caption{CLAP$_\text{score}$}
    \end{subfigure}
    \hfill
    \begin{subfigure}[t]{0.32\textwidth}
        \centering
        \begin{tikzpicture}
        \begin{axis}[
            width=1.2\textwidth,
            height=1.0\textwidth,
            xlabel={Iteration},
            xtick={1,2,3,4,5},
            ylabel={},
            grid=both,
            grid style={dotted, gray!50},
            mark options={solid},
            ticklabel style={font=\small}
        ]
        \addplot[color=blue, thick, mark=o] coordinates {
            (1, 78.2) (2, 77.0) (3, 75.8) (4, 75.5) (5, 75.9)
        };
        \addplot[color=red, thick, mark=o] coordinates {
            (1, 79.1) (2, 77.5) (3, 75.0) (4, 76.0) (5, 77.6)
        };
        \end{axis}
        \end{tikzpicture}
        \caption{FD$_\text{openl3}$}
    \end{subfigure}
    \hfill
    \begin{subfigure}[t]{0.32\textwidth}
        \centering
        \begin{tikzpicture}
        \begin{axis}[
    width=1.2\textwidth,
    height=1.0\textwidth,
    xlabel={Iteration},
    xtick={1,2,3,4,5},
    ylabel={},
    grid=both,
    grid style={dotted, gray!50},
    mark options={solid},
    ticklabel style={font=\small},
    legend style={font=\small, at={(0.05,0.95)}, anchor=north west},
]
\addplot[color=blue, thick, mark=o] coordinates {
    (1, 1.17) (2, 1.18) (3, 1.21) (4, 1.25) (5, 1.38)
};
\addplot[color=red, thick, mark=o] coordinates {
    (1, 1.18) (2, 1.19) (3, 1.19) (4, 1.24) (5, 1.35)
};
\legend{$\mathcal{L}_{\text{DPO-FM}}$, $\mathcal{L}_{\text{CRPO}}$}
\end{axis}
        \end{tikzpicture}
        \caption{KL$_\text{passt}$}
    \end{subfigure}

    \caption{Comparison between $\mathcal{L}_{\text{DPO-FM}}$ and $\mathcal{L}_{\text{CRPO}}$ w.r.t. (a) CLAP$_\text{score}$, (b) FD$_\text{openl3}$, and (c) KL$_\text{passt}$ across iterations.}
    \label{fig:crpovsdpo}
\end{figure}

As shown in \cref{fig:crpo_dpo}, the winning and losing losses of both $\mathcal{L}_{\text{CRPO}}$ and $\mathcal{L}_{\text{DPO-FM}}$ increase with each iteration, along with their difference/margin. Despite the increase in losses, \cref{fig:crpovsdpo} shows that benchmark performance improves, with $\mathcal{L}_{\text{CRPO}}$ achieving superior results in CLAP$_{\text{score}}$ while maintaining similar KL${_\text{passt}}$ and FD$_{\text{openl3}}$ across all iterations. We observe a notable acceleration in loss growth from $\mathcal{L}_{\text{DPO-FM}}$ after iteration 3, which may indicate performance saturation or degradation. In contrast, $\mathcal{L}_{\text{CRPO}}$ exhibits a more gradual and stable increase in loss, maintaining a smaller margin and more controlled growth, leading to less performance degradation as compared to $\mathcal{L}_{\text{DPO-FM}}$. This highlights the role of the \textit{winning loss} as a regularizer of the optimization dynamics by preventing the increase in margin at the cost of unmitigated increase of both \textit{winning loss} and \textit{losing loss}.


Our findings of increase in winning and losing losses in tandem with the margin is consistent with aligning LLMs with DPO~\cite{rafailov2024rqlanguagemodel}.  This paradoxical performance improvement from both $\mathcal{L}_{\text{CRPO}}$ and $\mathcal{L}_{\text{DPO-FM}}$ is also noted by \citet{rafailov2024scalinglawsrewardmodel} in the context of LLMs.

\begin{tcolorbox}[colback=blue!5!white, colframe=NotSureBlue, 
                  title=\textbf{TL;DR}, 
                  fonttitle=\bfseries, 
                  coltitle=white, 
                  boxrule=0.5mm, 
                  arc=2mm, 
                  left=2mm, 
                  right=2mm,
                  top=2mm,
                  bottom=2mm]

\begin{enumerate}[label=\textbf{\arabic*.}, leftmargin=*, wide]
    \item \textbf{Model Comparison}: 
    \begin{itemize}
        \item \model{} outperforms prior works in almost all objective metrics on \textit{AudioCaps}, especially for prompts with multiple events.
        \item It achieves superior performance in FD$_{\text{openl3}}$, CLAP$_{\text{score}}$, and Inception Score (IS), with notable efficiency gains (lowest inference time).
        \item Only \texttt{Tango 2} marginally surpasses \model{} in KL$_{\text{passt}}$.
    \end{itemize}

    
    \item \textbf{Multi-Event Prompts}:
    \begin{itemize}
        \item The margin in CLAP$_{\text{score}}$ between \model{} and baselines is larger for multi-event inputs, demonstrating its capability to handle complex and nuanced scenarios.
    \end{itemize}

    \item \textbf{Training Strategies:}
    \begin{itemize}
        \item Online batched data generation significantly outperforms offline strategies, preventing performance degradation caused by over-optimization.
        \item Online training maintains consistent improvement across CLAP$_{\text{score}}$ and KL$_{\text{passt}}$ over iterations.
    \end{itemize}

    \item \textbf{Preference Optimization:}
    \begin{itemize}
        \item \method{} dataset leads to better results than other preference datasets like BATON and Audio-Alpaca across all metrics.
        \item Larger $N$ in the Best-of-N policy enhances CLAP$_{\text{score}}$ and KL$_{\text{passt}}$, validating CLAP as an effective reward model.
    \end{itemize}

    \item \textbf{Optimization Techniques}:
    \begin{itemize}
        \item $\mathcal{L}_{\text{CRPO}}$ demonstrates more stable and effective optimization than $\mathcal{L}_{\text{DPO-FM}}$, with reduced performance saturation and better benchmark results.
        \item The controlled growth in optimization metrics with $\mathcal{L}_{\text{CRPO}}$ highlights its robustness for rectified training processes.
    \end{itemize}

    \item  \textbf{Inference Time}:
    \begin{itemize}
        \item While delivering superior performance, \model{} also boasts a much lower inference time, resulting in greater efficiency compared to other models.
        \item \model{} shows less performance decline compared to other models when sampling at fewer steps.
    \end{itemize}
\end{enumerate}
\end{tcolorbox}

\subsection{Human Evaluation Results}

The results of the human evaluation are presented in \cref{tab:human_eval}, with detailed comparisons of the models across the evaluated metrics: z-scores, rankings, and Elo scores for both overall audio quality (OVL) and relevance to the text input (REL).

\textbf{z-scores:} z-score mitigates individual scoring biases by normalization into a standard normal variable with zero mean and one standard deviation. \model{} demonstrated the highest performance across both metrics, with z-scores of 0.2486 for OVL and 0.6919 for REL. This indicates its superior quality and strong alignment with the input prompts. Conversely, \texttt{AudioLDM 2} scored the lowest with z-scores of -0.3020 (OVL) and -0.4936 (REL), suggesting both lower sound quality and weaker adherence to textual inputs as compared to the other models.

\textbf{Ranking:} Rankings provide an ordinal measure of performance, complementing z-score findings. \model{} achieved the best rankings with a mean rank of 1.7 (OVL) and 1.1 (REL), and mode ranks of 2 (OVL) and 1 (REL), affirming its superiority in subjective evaluations. In contrast, \texttt{AudioLDM 2} consistently ranked lowest, with mean ranks of 3.5 (OVL) and 3.7 (REL), and mode ranks of 4 for both metrics. \texttt{StableAudio} and \texttt{Tango 2} had similar mean ranks for OVL (2.4), but \texttt{Tango 2} outperformed \texttt{StableAudio} on REL (mean ranks: 1.9 vs. 3.3). Notably, \texttt{StableAudio}'s bimodal OVL ranks (modes 1 and 3) suggest polarized annotator perceptions, likely due to misalignment between prompts and outputs, as reflected in its REL rankings (mean 3.3, mode 3).

\textbf{Elo Scores:} Elo scores provide a probabilistic measure of model performance, by accounting for pairwise relative performance. Here, \model{} again excelled, achieving the highest Elo scores for both OVL (1,501) and REL (1,628). The Elo results highlight the robustness of \model{}, as it consistently outperformed other models in pairwise comparisons. \texttt{Tango 2} emerged as the second-best performer, with Elo scores of 1,419 (OVL) and 1,507 (REL). \texttt{StableAudio} follows, showing competitive performance in OVL (1,444), but a weaker REL score (1,268). Like other metrics, \texttt{AudioLDM 2} ranked last with the least Elo scores (1,236 for OVL and 1,196 for REL).

\begin{tcolorbox}[colback=blue!5!white, colframe=NotSureBlue, 
                  title=\textbf{TL;DR}, 
                  fonttitle=\bfseries, 
                  coltitle=white, 
                  boxrule=0.5mm, 
                  arc=2mm, 
                  left=2mm, 
                  right=2mm,
                  top=2mm,
                  bottom=2mm]

\begin{enumerate}[label=\textbf{\arabic*.}, leftmargin=*, wide]
    \item \model{} consistently demonstrated superior performance across all metrics, highlighting its strength in generating high-quality, text-relevant audio. This is particularly evident in its significant lead in the REL metrics, showcasing its robust capability to align with complex, multi-event prompts.
    
    \item \texttt{Tango 2} performed strongly in REL, reflecting its alignment capability. However, it slightly lagged behind \texttt{TangoFlux} in OVL, indicating potential room for improvement in audio clarity and naturalness.
    
    \item \texttt{Stable Audio Open} displayed competitive performance in OVL, but its REL scores suggest limitations in accurately and faithfully representing complex text inputs.
    
    \item \texttt{AudioLDM2} consistently underperformed across all metrics, reflecting challenges in both audio quality and relevance to complex prompts. This positions it as the least preferred model in this evaluation.
\end{enumerate}

\end{tcolorbox}

\begin{table}[ht!]
\centering
\resizebox{\linewidth}{!}{
\begin{tabular}{l*{8}{c}}
\toprule
\multirow{4}{*}{\textbf{Model}} & \multicolumn{2}{c}{\textbf{z-scores}} & \multicolumn{4}{c}{\textbf{Ranking}} & \multicolumn{2}{c}{\textbf{Elo}} \\
\cmidrule(lr){2-3}\cmidrule(lr){4-7}\cmidrule(lr){8-9}
& \textbf{OVL} & \textbf{REL} & \multicolumn{2}{c}{\textbf{OVL}} & \multicolumn{2}{c}{\textbf{REL}} & \textbf{OVL} & \textbf{REL}\\
\cmidrule(lr){4-5}\cmidrule(lr){6-7}
&&& \textbf{Mean} & \textbf{Mode} & \textbf{Mean} & \textbf{Mode} & & \\
\midrule
\texttt{AudioLDM 2} & -0.3020 & -0.4936 & 3.5 &  4 & 3.7  & 4 & 1,236 & 1,196 \\
\texttt{SA Open} & 0.0723 & -0.3584 & 2.4  & 1, 3 & 3.3  & 3 & 1,444 & 1,268 \\
\texttt{Tango 2} & -0.019 & 0.1602 & 2.4 & 2 & 1.9 & 2 & 1,419 & 1,507 \\
\rowcolor{pink!20}
\model{} & \textbf{0.2486} & \textbf{0.6919} & \textbf{1.7}  & \textbf{2} &\textbf{ 1.1}  & \textbf{1} & \textbf{1,501} & \textbf{1,628} \\
\bottomrule
\end{tabular}
}
\caption{Human evaluation results on OVL (quality) and REL (relevance); \texttt{SA Open} $\coloneqq$ \texttt{Stable Audio Open}.
}
\label{tab:human_eval}
\end{table}

\subsection{Inference Time vs Performance}

\model{} beats the other models in terms of performance per unit of inference time, measured w.r.t. CLAP and FD score. See \cref{app:it-vs-p} for more details.

\section{Related Works}
\textbf{Text-To-Audio Generation.} TTA Generation has lately drawn attention due to AudioLDM \cite{liu2024audioldm2learningholistic,liu2023audioldmtexttoaudiogenerationlatent}, Tango \cite{majumder2024tango2aligningdiffusionbased,ghosal2023texttoaudiogenerationusinginstructiontuned,kong2024improvingtexttoaudiomodelssynthetic}, and Stable Audio \cite{evans2024fasttimingconditionedlatentaudio,evans2024stableaudioopen,evans2024longformmusicgenerationlatent} series of models. These adopt the diffusion framework \cite{song2020generativemodelingestimatinggradients,rombach2022highresolutionimagesynthesislatent,song2022denoisingdiffusionimplicitmodels,ho2020denoisingdiffusionprobabilisticmodels}, which trains a latent diffusion model conditioned on textual embedding. Another common framework for TTA generation is flow matching which was  employed in models such as VoiceBox \cite{le2023voiceboxtextguidedmultilingualuniversal}, AudioBox \cite{vyas2023audioboxunifiedaudiogeneration}, FlashAudio \cite{liu2024flashaudiorectifiedflowsfast}.

\textbf{Alignment Method.} Preference optimization is the standard approach for aligning LLMs, achieved either by training a reward model to capture human preferences \cite{ouyang2022traininglanguagemodelsfollow} or by using the LLM itself as the reward model \cite{rafailov2024directpreferenceoptimizationlanguage}. Recent advances improve this process through iterative alignment, leveraging human annotators to construct preference pairs or utilizing pre-trained reward models. \cite{kim2024sdpodontusedata,chen2024selfplayfinetuningconvertsweak,gulcehre2023reinforcedselftrainingrestlanguage,yuan2024selfrewardinglanguagemodels}. Verifiable answers can enhance the construction of preference pairs. For diffusion and flow-based models, Diffusion-DPO shows that these models can be aligned similarly \cite{wallace2023diffusionmodelalignmentusing}. However, constructing preference pairs for TTA is challenging due to the absence of "gold" audio for given text prompts and the subjective nature of audio. BATON \cite{liao2024batonaligningtexttoaudiomodel} relies on human annotations, which is not scalable.

\section{Conclusion}

We introduce \model{}, a fast flow-based text-to-audio model aligned using synthetic preference data generated online during training. Objective and human evaluations show that \model{} produces audio more representative of user prompts than existing diffusion-based models, achieving state-of-the-art performance with significantly fewer parameters. Additionally, \model{} demonstrates greater robustness, maintaining performance even when sampling with fewer time steps. These advancements make \model{} a practical and scalable solution for widespread adoption.
\nocite{langley00}

\bibliography{iclr2024_conference}
\bibliographystyle{iclr2024_conference.bst}

\newpage
\appendix
\onecolumn
\section{Appendix}
\subsection{Effect of CFG scale}

We conduct an ablation of the effect of CFG scale for \model{} and show the result in Table \ref{tab:cfg}.  It reveals a trade-off: higher CFG values improve FD score (lower FD) but slightly reduce semantic alignment (CLAP score), which peaks at CFG=3.5. The results emphasize CFG=3.5 as the optimal balance between fidelity and semantic relevance.
\label{app:cfg}
\begin{table*}[t]
\centering
\begin{tabular}{@{}lccccccc@{}}
\toprule
Model & Steps &CFG  & FD$_{\text{openl3}}$ $\downarrow$ & KL$_{\text{passt}}$ $\downarrow$ & CLAP$_{\text{score}}$ $\uparrow$ \\ \midrule                        

\multirow{4}{*}{\textbf{\model{}}}  
 & 50 &3.0& 77.7& \textbf{1.14}  & 0.479   \\
& 50 &3.5& 76.1& \textbf{1.14}  & \textbf{0.481}   \\
  & 50 &4.0& 74.9& 1.15  & 0.476   \\
    & 50 &4.5& 75.1& 1.15  & 0.480  \\
  & 50 &5.0& \textbf{74.6}& 1.15  & 0.472  \\

\bottomrule 
\end{tabular}
\caption{\model{} with different classifier free guidance (CFG) values.}
\label{tab:cfg}

\end{table*}

\subsection{Inference Time vs Performance Comparison}
\label{app:it-vs-p}

\begin{figure*}[htb] 
    \centering
    \begin{subfigure}[b]{0.49\textwidth}
        \centering
        \includegraphics[width=\textwidth]{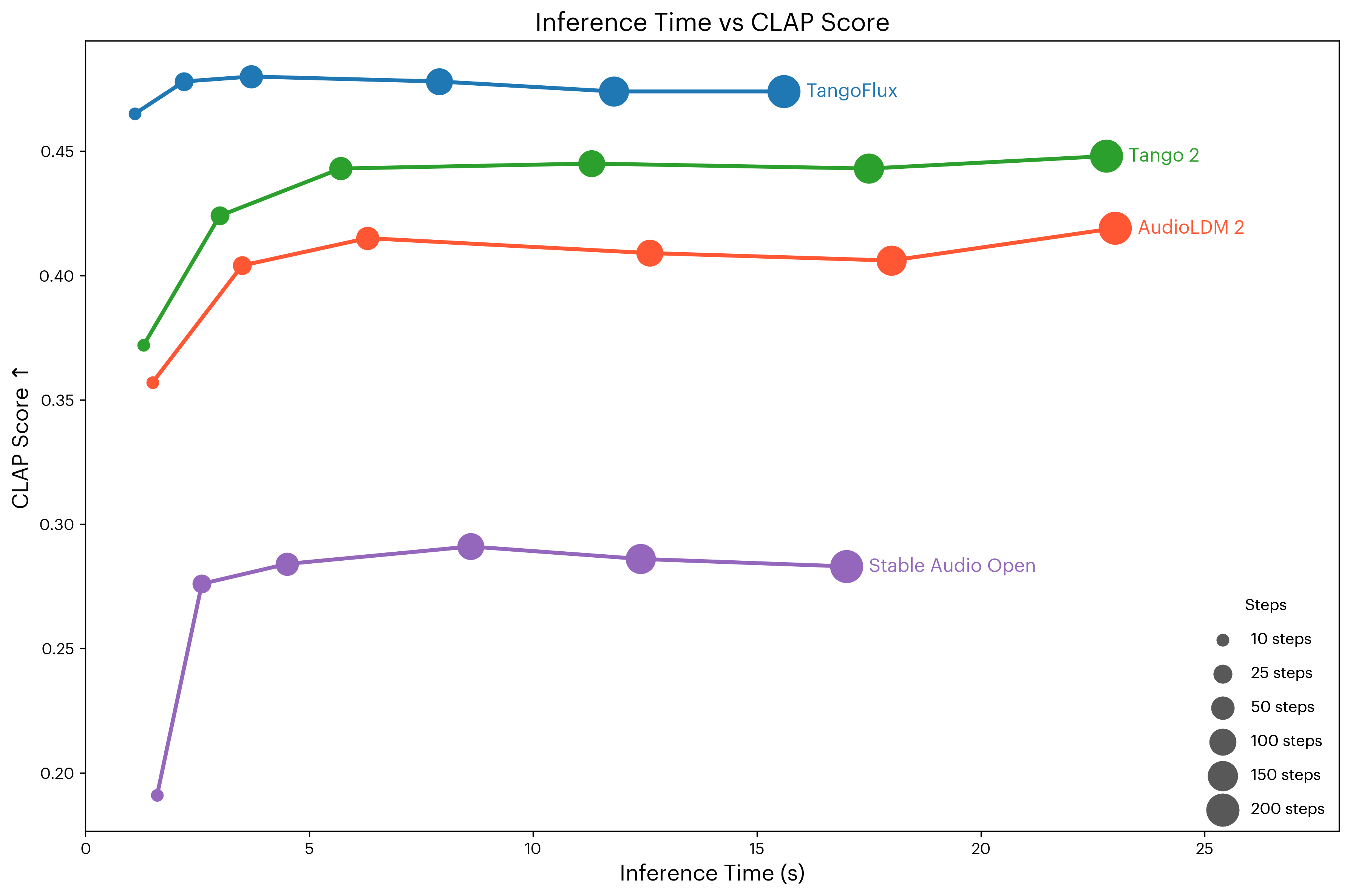}
        \caption{}
        \label{fig:clap-vs-time}
    \end{subfigure}
    \hfill
    \begin{subfigure}[b]{0.49\textwidth}
        \centering
        \includegraphics[width=\textwidth]{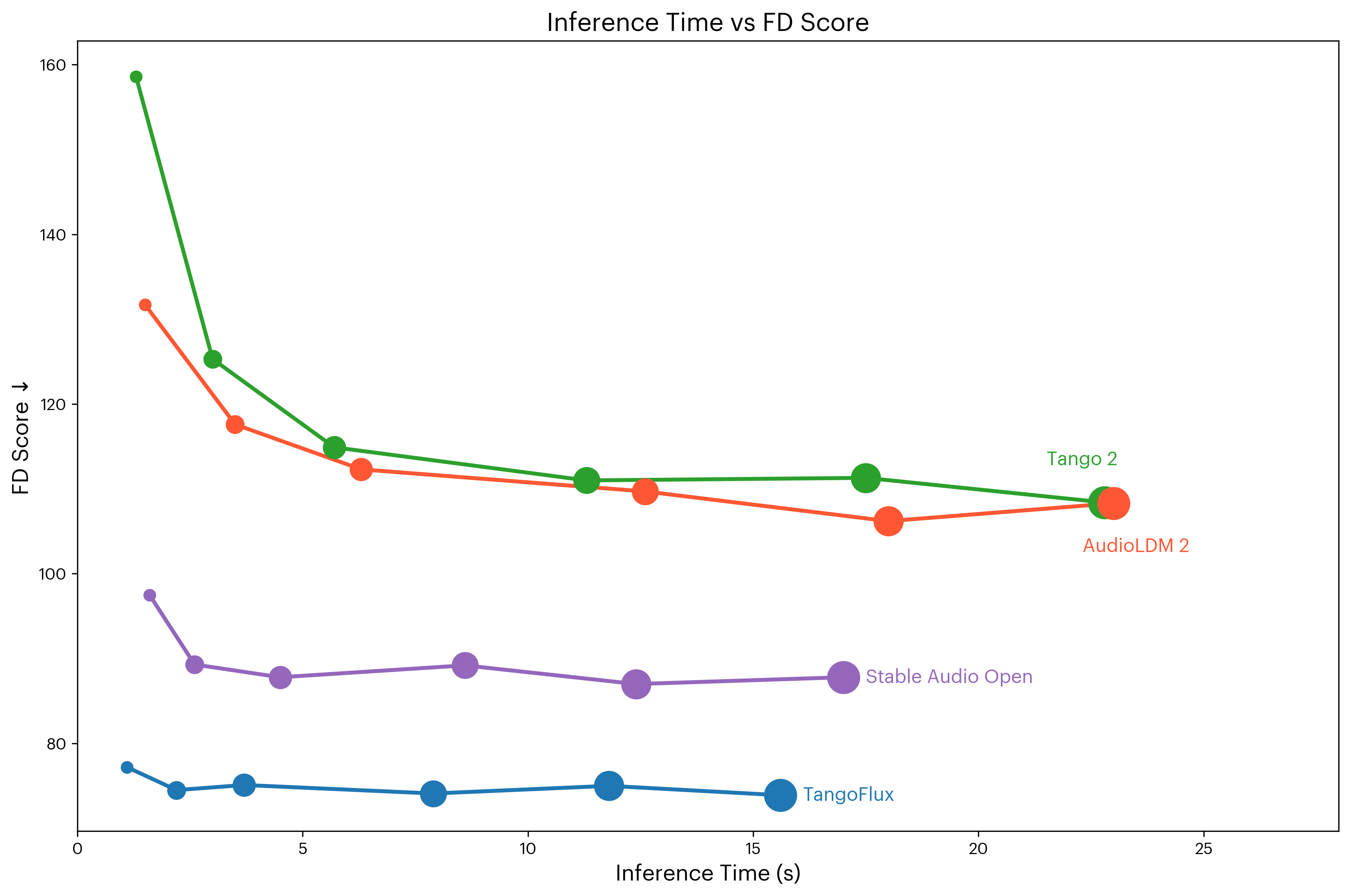}
        \caption{}
        \label{fig:fd-vs-time}
    \end{subfigure}

    \caption{Comparison of (a) CLAP and (b) FD Scores vs Inference Time for each model. Results are plotted for step counts of 10, 25, 50, 100, 150, and 200.}
    \label{fig:inference-time}
\end{figure*}

Across models, we compare the trajectory of CLAP and FD scores with increasing inference time for steps 10, 25, 50, 100, 150, and 200, as shown in \Cref{fig:inference-time}. \model{} demonstrates a remarkable balance between efficiency and performance, consistently achieving higher CLAP scores and lower FD scores while requiring significantly less inference time compared to other models. For example, at 50 steps, \model{} achieves a CLAP score of 0.480 and an FD score of 75.1 in just 3.7 seconds. In comparison, \texttt{Stable Audio Open} requires 4.5 seconds for the same step count but only achieves a CLAP score of 0.284 (41\% lower than \model{}) and an FD score of 87.8 (17\% worse than \model{}). This demonstrates that \model{} achieves superior performance metrics in less time. Additionally, at a lower step count of 10, \model{} maintains strong performance with a CLAP score of 0.465 and an FD score of 77.2 in just 1.1 seconds. In contrast, \texttt{Audioldm2} at the same step count achieves a lower CLAP score of 0.357 (23\% lower) and a significantly worse FD score of 131.7 (70\% higher), while requiring 1.5 seconds (36\% more time). We also observe that reducing the step count from 200 to 10 has a minimal impact on \model{}'s performance, highlighting its robustness. Specifically, \model{}'s CLAP score decreases by only 3.2\% (from 0.480 to 0.465), and its FD score increases by only 4.5\% (from 73.9 to 77.2). In contrast, \texttt{Tango 2} shows a larger degradation, with its CLAP score decreasing by 16.0\% (from 0.443 to 0.372) and its FD score increasing by 37.8\% (from 108.4 to 158.6).

These results highlight \model{}'s effectiveness in delivering high-quality outputs with lower computational requirements, making it a highly efficient choice for scenarios where inference time is critical.

\subsection{Human Evaluation}
\label{app:ood}

\begin{figure}[h]
    \centering
    \includegraphics[width=\linewidth]{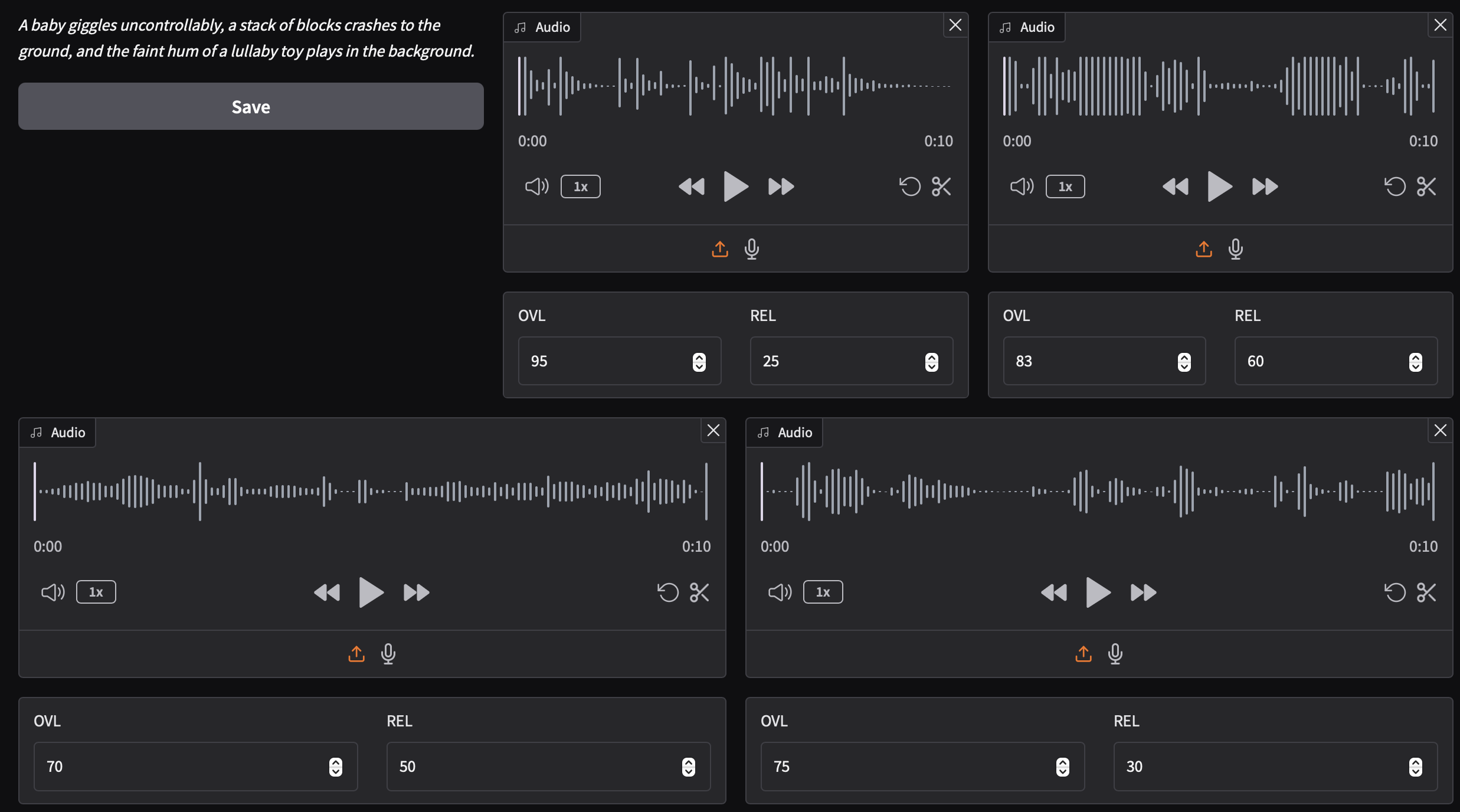}
    \caption{The Gradio-based human evaluation form created for the annotators to score the model generated audios with respect to the input prompts.}
    \label{fig:heval-form}
\end{figure}

The human evaluation was performed using a web-based Gradio\footnote{https://www.gradio.app} app. Each annotator was presented with 20 prompts, each having four audio samples generated by four distinct text-to-audio models, shuffled randomly, as shown in \cref{fig:heval-form}. Before the annotation process, the annotators were instructed with the following directive:
\begin{mdframed}[backgroundcolor=green!5]
Welcome \emph{username}

\noindent\# \textbf{\Large Instructions for evaluating audio clips}

\textbf{Please carefully read the instructions below}.

\noindent\#\# \textbf{\large Task}

You are to evaluate four 10-second-long audio outputs to each of the 20 prompts below. These four outputs are from
four different models. You are to judge each output with respect to two qualities:
\begin{itemize}
    \item Overall Quality (OVL): The overall quality of the audio is to be judged on a scale from 0 to 100: 0 being absolute noise with no discernible feature. Whereas, 100 is perfect.
    \textbf{Overall fidelity, clarity, and noisiness of the audio are important here.}
    \item Relevance (REL): The extent of audio alignment with the prompt is to be judged on a scale from 0 to 100: with 0 being absolute irrelevance to the input description. Whereas, 100 is a perfect representation of the input description.
    \textbf{You are to judge if the concepts from the input prompt appear in the audio in the described temporal order.}
\end{itemize}

\textbf{You may want to compare the audios of the same prompt with each other during the evaluation.}
                
\noindent\#\# \textbf{\large Listening guide}

\begin{enumerate}
\item Please use a head/earphone to listen to minimize exposure to the external noise.
\item Please move to a quiet place as well, if possible.
\end{enumerate}

\noindent\#\# \textbf{\large UI guide}

\begin{enumerate}
\item Each audio clip has two attributes OVL and REL below. You may select the appropriate option from the dropdown list.
\item To save your judgments, please click on any of the \emph{save} buttons. All the \emph{save} buttons function identically. They are placed everywhere to avoid the need to scroll to save.
\end{enumerate}

Hope the instructions were clear. Please feel free to reach out to us for any queries.

\end{mdframed}

\subsubsection{Evaluation Dataset}
\label{app:heval-ds}

To evaluate the instruction-following capabilities and robustness of TTA models, we created 50 out-of-distribution complex captions, such as ``\textit{A pile of coins spills onto a wooden table with a metallic clatter, followed by the hushed murmur of a tavern crowd and the creak of a swinging door}''. These captions describe 3--6 events and aim to go beyond conventional or overused sounds in the evaluation sets, such as simple animal noises, footsteps, or city ambiance. Events were identified using GPT4o to evaluate the captions generated. Each of the generated prompts contains multiple events including several where the temporal order of the events must be maintained. Details of our caption generation template and samples of generated captions can be found in the \cref{app:ood}.

\subsubsection{Metrics}
\label{app:heval-metrics}

We report three key metrics for subjective evaluation:

\textbf{$z$-score:} The average of the scores assigned by individual annotators. Due to the subjective nature of these scores and the significant variance observed in the annotator scoring patterns, the ratings were normalized to z-scores at the annotator level:
\(
z_{ij} = (s_{ij} - \mu_{i})/{\sigma_{i}}
\). $z_{ij}$: The z-score for annotator $i$'s score of model $M_j$. This is the score after applying z-score normalization. $s_{ij}$: The raw score assigned by annotator $i$ to model $j$. This is the original score before normalization. $\mu_{i}$: The mean score assigned by annotator $i$ across all models. It represents the central tendency of the annotator’s scoring pattern. $\sigma_{i}$: The standard deviation of annotator $i$'s scores across all models. This measures the variability or spread in the annotator's ratings.

This normalization procedure adjusts the raw scores, centering them around the annotator's mean score and scaling by the annotator's score spread (standard deviation). This ensures that scores from different annotators are comparable, helping to mitigate individual scoring biases.

 \textbf{Ranking:} Despite z-score normalization, the variability in annotator scoring can still introduce noise into the evaluation process. To address this, models are also ranked based on their absolute scores. We utilize the mean (average rank of a model), and mode (the most common rank of a model) as metrics for evaluating these rankings.
 
 \textbf{Elo:} Elo-based evaluation, a widely adopted method in language model assessment, involves pairwise model comparisons. We first normalized the absolute scores of the models using z-score normalization and then derived Elo scores from these pairwise comparisons. Elo score mitigates the noise and inconsistencies observed in scoring and ranking techniques. Specifically, Elo considers the relative performance between models rather than relying solely on absolute or averaged scores, providing a more robust measure of model quality under subjective evaluation. While ranking-based evaluation provides an ordinal comparison of models, determining the order of performance (e.g., Model A ranks first, Model B ranks second), it does not capture the magnitude of differences between ranks. For instance, if the difference between the first and second rankers is minimal, this is not evident from ranks alone. Elo scoring addresses this limitation by integrating both ranking and pairwise performance data. In ranking-based systems, the rank \(R_i\) of a model \(M_i\) is determined purely by its position relative to others:
\[
\scriptsize
R_i = \text{position of } M_i \text{ in the sorted list of models based on performance.}
\] However, this approach fails to quantify: 1) The gap in performance between consecutive ranks. 2) The consistency of relative performance across different pairwise comparisons. Elo scoring provides a probabilistic measure of model performance based on pairwise comparisons. By leveraging annotator scores, Elo assigns a continuous score \(E_i\) to each model \(M_i\), capturing its relative strength.

\subsubsection{Prompts Used in the Evaluation}
\renewcommand{\arraystretch}{1.5}
\rowcolors{2}{gray!15}{white}
\begin{longtable}{|p{7.5cm}|c|c|}
\hline
\rowcolor[HTML]{CFE2F3} 
\textbf{Prompts} & \textbf{Multiple Events} & \textbf{Temporal Events} \\ \hline
A robotic arm whirs frantically while an electric plasma arc crackles and a metallic voice counts down ominously, interspersed with glass vials clinking to the floor. & \greentick{} & \greentick{} \\ \hline
Unfamiliar chirps overlap with a low, throbbing hum as bioluminescent plants audibly crackle and squelch with movement. & \greentick{} & \redcross{} \\ \hline
Dripping water echoes sharply, a distant growl reverberates through the cavern, and soft scraping metal suggests something lurking unseen. & \greentick{} & \redcross{} \\ \hline
Alarms blare with rising urgency as fragments clatter against a metallic hull, interrupted by a faint hiss of escaping air. & \greentick{} & \greentick{} \\ \hline
Hundreds of tiny wings buzz with a chaotic pitch shift, joined by the faint clattering of mandibles and an organic squish as they collide. & \greentick{} & \redcross{} \\ \hline
Jagged rocks crumble underfoot while distant ocean waves crash below, punctuated by the sudden snap of a rope. & \greentick{} & \greentick{} \\ \hline
Digital beeps and chirps meld with overlapping chatter in multiple languages, as automated drones whiz past, scanning barcodes audibly. & \greentick{} & \redcross{} \\ \hline
Rusted swings creak in rhythmic disarray, a faint mechanical laugh stutters from a distant speaker, and the sound of gravel crunches under unseen footsteps. & \greentick{} & \redcross{} \\ \hline
Bubbling lava gurgles ominously, instruments beep irregularly, and faint crackling signals static from a failing radio. & \greentick{} & \greentick{} \\ \hline
Tiny pops and hisses of chemical reactions intermingle with the rhythmic pumping of a centrifuge and the soft whirr of air filtration. & \greentick{} & \redcross{} \\ \hline
The faint hiss of a gas leak grows louder as metal chains rattle and a single marble rolls across the floor. & \greentick{} & \greentick{} \\ \hline
A hand slaps a table sharply, followed by the shuffle of playing cards and the hum of an overhead fan. & \greentick{} & \greentick{} \\ \hline
A train horn blares in the distance as a bicycle bell chimes and a soda can pops open with a fizzy hiss. & \greentick{} & \redcross{} \\ \hline
A drawer creaks open, papers rustle wildly, and the sharp click of a lock snapping shut echoes. & \greentick{} & \redcross{} \\ \hline
A burst of static interrupts soft typing sounds, followed by the distant chirp of a pager and a cough. & \greentick{} & \greentick{} \\ \hline
A heavy book thuds onto a desk, accompanied by the faint buzz of a fluorescent light and a muffled sneeze. & \greentick{} & \redcross{} \\ \hline
The sharp squeak of sneakers on a gym floor blends with the rhythmic bounce of a basketball and the screech of a metal door. & \greentick{} & \redcross{} \\ \hline
An elevator dings, its doors sliding open, as muffled voices overlap with the shuffle of heavy bags. & \greentick{} & \redcross{} \\ \hline
A clock ticks steadily, a light switch clicks on, and the crackle of a fire igniting briefly fills the silence. & \greentick{} & \greentick{} \\ \hline
A fork scrapes a plate, water drips slowly into a sink, and the faint hum of a refrigerator lingers in the background. & \greentick{} & \redcross{} \\ \hline
A cat hisses sharply as glass shatters nearby, followed by hurried footsteps and the slam of a closing door. & \greentick{} & \greentick{} \\ \hline
A parade marches through a town square, with drumbeats pounding, children clapping, and a horse neighing amidst the commotion. & \greentick{} & \greentick{} \\ \hline
A basketball bounces rhythmically on a court, shoes squeak against the floor, and a referee’s whistle cuts through the air. & \greentick{} & \redcross{} \\ \hline
A baby giggles uncontrollably, a stack of blocks crashes to the ground, and the faint hum of a lullaby toy plays in the background. & \greentick{} & \redcross{} \\ \hline
The rumble of a subway train grows louder, followed by the screech of brakes and muffled announcements over a crackling speaker. & \greentick{} & \greentick{} \\ \hline
A beekeeper moves carefully as bees buzz intensely, a smoker puffs softly, and wooden frames creak as they’re lifted. & \greentick{} & \redcross{} \\ \hline
A dog shakes off water with a noisy splatter, a bicycle bell rings, and a distant lawnmower hums faintly in the background. & \greentick{} & \redcross{} \\ \hline
Books fall off a shelf with a heavy thud, a chair scrapes loudly across a wooden floor, and a surprised gasp echoes. & \greentick{} & \redcross{} \\ \hline
A soccer ball hits a goalpost with a metallic clang, followed by cheers, clapping, and the distant hum of a commentator’s voice. & \greentick{} & \greentick{} \\ \hline
A hiker’s pole taps against rocks, a mountain goat bleats sharply, and loose gravel tumbles noisily down a steep slope. & \greentick{} & \greentick{} \\ \hline
A rooster crows loudly at dawn, joined by the rustle of feathers and the crunch of chicken feed scattered on the ground. & \greentick{} & \redcross{} \\ \hline
A carpenter saws through wood with steady strokes, a hammer strikes nails rhythmically, and a measuring tape snaps back with a metallic zing. & \greentick{} & \redcross{} \\ \hline
A frog splashes into a pond as dragonflies buzz nearby, accompanied by the distant croak of toads echoing through the marsh. & \greentick{} & \redcross{} \\ \hline
The crack of a whip startles a herd of cattle, their hooves clatter against a dirt path as a rancher shouts commands. & \greentick{} & \redcross{} \\ \hline
A paper shredder whirs noisily, the rustle of documents being fed in grows louder, and a stapler clicks shut in rapid succession. & \greentick{} & \redcross{} \\ \hline
An elephant trumpets in the savanna as a herd stomps through dry grass, accompanied by the buzz of flies and the distant roar of a lion. & \greentick{} & \redcross{} \\ \hline
A mime claps silently as a juggling act clinks glass balls together, and a crowd bursts into laughter at the clatter of a dropped prop. & \greentick{} & \redcross{} \\ \hline
A train conductor blows a sharp whistle, metal wheels screech on the rails, and passengers murmur while settling into their seats. & \greentick{} & \greentick{} \\ \hline
A squirrel chitters nervously as acorns drop from a tree, landing with dull thuds, while leaves rustle above in quick bursts of movement. & \greentick{} & \redcross{} \\ \hline
A blacksmith hammers molten iron with rhythmic clangs, a bellows pumps air with a whoosh, and sparks sizzle on a stone floor. & \greentick{} & \redcross{} \\ \hline
A skateboard grinds loudly against a metal rail, followed by the sharp slap of wheels hitting pavement and a triumphant cheer from the rider. & \greentick{} & \redcross{} \\ \hline
An old typewriter clacks rapidly as paper rustles with each keystroke, interrupted by the sharp ding of the carriage return. & \greentick{} & \redcross{} \\ \hline
A pack of wolves howls in unison as dry leaves crunch underfoot, and the faint trickle of a nearby stream echoes through the forest. & \greentick{} & \redcross{} \\ \hline
\caption{Prompts used in human evaluation and their characteristics.}
\end{longtable}

\subsection{BATON as a Preference Dataset}
\label{app:baton}

BATON contains human-annotated data where annotators assign a binary label of 0 or 1 to each audio sample based on its alignment with a given prompt: 1 indicates alignment, while 0 indicates misalignment. We construct a preference dataset by pairing audio samples labeled 1 (winners) with those labeled 0 (losers) for the same prompt, creating a set of winner-loser pairs.

\subsection{Multi-Staged Relation-Aware Evaluation}
\begin{figure}[t]
\begin{minipage}[t]{.5\linewidth}
\centering
\small
    \renewcommand{\arraystretch}{0.5}
    \begin{tabular}{m{1.2cm}|m{1.4cm}|m{3.5cm}}
    \hline
    \makecell[c]{Main\\ Relation} & \makecell[c]{Sub-\\Relation} & \makecell[c]{Sample Text Prompt} \\
    \hline
       \makecell[c]{Temporal \\ Order}  &  \makecell[c]{before;\\ after;\\ simultaneity} & \makecell[c]{generate dog barking audio,\\ followed by cat meowing;}\\
       \hline
       \makecell[c]{Spatial \\Distance} & \makecell[c]{close first;\\far first;\\ equal dist.} & \makecell[c]{generate dog barking audio\\ that is 1 meter away, follow-\\ed by another 5 meters away.}\\
       \hline
      \makecell[c]{Count}& \makecell[c]{count} & \makecell[c]{produce 3 audios: dog bark-\\ing, cat meowing and talking.}\\
       \hline
       \makecell{Composit\\ionality} & \makecell[c]{and; or;\\ not;\\ if-then-else} & \makecell[c]{create dog barking audio\\ or cat meowing audio.}\\
       \hline
    \end{tabular}
    \vspace{-1mm}
    \captionof{table}{\small Audio Events Relation Corpus.}
    \label{tab:relation_corpus}
\end{minipage}\quad
\hspace{2mm}
\begin{minipage}[t]{.4\linewidth}
\centering
\small
    \begin{tabular}{m{1.7cm}|m{4.cm}}
    \hline
    \makecell[c]{Main\\Category} & \makecell[c]{Sub-Category} \\ 
    \hline
       \makecell[c]{Human\\ Audio}  &  baby crying; talking; laughing; coughing; whistling \\
       \hline
       \makecell[c]{Animal\\ Audio} & cat meowing; bird chirping; dog barking; rooster crowing; sheep bleating\\
       \hline
       \makecell[c]{Machinery} & boat horn; car horn; door bell; paper shredder; telephone ring\\
       \hline
       \makecell[c]{Human-Object\\ Interaction} &vegetable chopping; door slam; footstep; keyboard typing; toilet flush\\
       \hline
        \makecell[c]{Object-Object\\ Interaction} & emergent brake; glass drop; hammer nailing; key jingling; wood sawing\\
       \hline
    \end{tabular}
    \vspace{-1mm}
    \captionof{table}{\small Audio Events Category Corpus.}
    \label{tab:event_corpus}
\end{minipage}
\label{fig:ritta}
\vspace{-4mm}
\end{figure}

\end{document}